\documentclass[aps,prb,preprint,showpacs,preprintnumbers,amsmath,amssymb]{revtex4}

\usepackage{graphicx}% Include figure files
\usepackage{dcolumn}% Align table columns on decimal point
\usepackage{bm}% bold math

\newcommand{\JPSJ}[3]{\textit{J. Phys. Soc. Jpn.}        \textbf{#1}, #2 (#3)}
\newcommand{\PRB}[3]{\textit{Phys. Rev. B}        \textbf{#1}, #2 (#3)}
\newcommand{\PR}[3]{\textit{Phys. Rev.}        \textbf{#1}, #2 (#3)}
\newcommand{\JPCM}[3]{\textit{J. Phys. Condens. Matter}        \textbf{#1}, #2 (#3)}

\newcommand{\PhB}[3]{\textit{Physica B}        \textbf{#1}, #2 (#3)}

\newcommand{\Scie}[3]{\textit{Science}        \textbf{#1}, #2 (#3)}
\newcommand{\NPh}[3]{\textit{Nature Physics}        \textbf{#1}, #2 (#3)}
\newcommand{\CPC}[3]{\textit{Computer Phys. Commun.}       \textbf{#1}, #2 (#3)}
\newcommand{\ZK}[3]{\textit{Zeit. Kristallogr.}       \textbf{#1}, #2 (#3)}

\newcommand{\SSC}[3]{\textit{Solid State Commun.}       \textbf{#1}, #2 (#3)}
\newcommand{\JPC}[3]{\textit{J. Phys. C: Solid St. Phys.}       \textbf{#1}, #2 (#3)}

\newcommand{\tn}{T_\mathrm{N}}

\newcommand{\rmi}{\mathrm{i}}   %for JPCM igai

\def\ffrac#1#2{{{#1}/{#2}}}

\def\expu#1{{  \mathrm{e}^{ {#1}  }  }}

\begin{document}

\preprint{}

\title{Theoretical study on the variation of ordering vector in 
Ce(Pd$_{1-x}$M$_x$)$_2$Al$_3$ (M$=${Ag, Cu})  }% Force line breaks with \\

\author{Kazunori Tanaka}
\email{ktanaka@ph.noda.tus.ac.jp}
\author{Takumi Araki}%
\author{Katsurou Hanzawa}%
\affiliation{%
Department of Physics, Faculty of Science and Technology,
Tokyo University of Science, Noda 278-8510, Japan \\
}%

\date{}% 

\begin{abstract}
In heavy-fermion  compounds, the crossover from the localized to 
itinerant heavy-fermion state is observed with lowering temperature,
frequently accompanied by magnetism.
Ordering vectors of magnetism often vary with applying pressure or with substituting atoms. 
In Ce(Pd$_{1-x}$M$_x$)$_2$Al$_3$ with $\mathrm{M}=\mathrm{Ag,Cu}$, 
the $(0,0,1/2)$-antiferromagnetic (AF), ferromagnetic (F), 
and another AF orders are observed for $x<0.05$, $0.1<x<0.4$, and 
$0.5<x$, respectively. 
This change in the ordering vector is considered to
be caused by the change in the conduction-band structures. 
Using the anisotropic RKKY interaction model reflecting the spacial anisotropic distribution of the $f$ states 
and also the conduction-band structures, % obtained by the band calculation, 
we study the change in the ordering vector of Ce(Pd$_{1-x}$M$_x$)$_2$Al$_3$ 
with $x$ theoretically. 
As a result, 
the variation of the ordering vector is explained 
by treating the substitution of atoms as the conduction-electron doping, 
and the ordering vector of the AF state for $x>0.5$ is considered to be $(1/2,0,1/2)$. 
\end{abstract}


\maketitle
\section{Introduction}
Heavy-fermion systems, such as Ce-based and U-based compounds, 
exhibit at low temperatures the magnetic orders with various ordering vectors, 
which often vary with applying pressure or with substituting atoms. 
In this paper, we treat the magnetism of
Ce(Pd$_{1-x}$M$_x$)$_2$Al$_3$  with $\mathrm{M}=\mathrm{Ag,Cu}$. 
CePd$_2$Al$_3$ crystallizes in the PrNi$_2$Al$_3$-type simple hexagonal (SH) structure, 
which has the $D_{6h}$ point-group and the P6/mmm space-group symmetry,
with the lattice constants approximately $a=5.471$ {\AA}  and $c=4.216$ {\AA}\cite{JPSJ61-1461}. 
The magnitude of the Sommerfeld coefficient 
($\gamma \sim 380$ mJ\,mol$^{-1}$K$^{-2}$)\cite{JPSJ61-1461} 
indicates that this compound is in the heavy-fermion state. 
The $f$ states with the total angular momentum  $J=5/2$, 
which mainly contribute to the formation of the heavy fermion state, 
are divided into three doublets with $M=\pm 1/2$, $\pm 3/2$, and $\pm5/2$, 
where $M$ is the $z$ ($c$)  component of $J$, in the hexagonal 
crystalline electric field (CEF). 
CePd$_2$Al$_3$ exhibits a large value of $\chi_a/\chi_c$ $(\sim 20)$\cite{PRB49-15759} 
in the paramagnetic phase, 
indicating that the CEF ground states for the $f$ electrons are composed of the $M=\pm 1/2$ states mainly. 
In fact, the susceptibility in the paramagnetic phase  is described well 
by assuming the  CEF ground states as $M=\pm 1/2$,  
and the excited states as $\pm 3/2$ (33 K) and $\pm 5/2$ (800 K)\cite{PRB49-15759}. 
CePd$_2$Al$_3$
orders in the antiferromagnetic (AF) state of  $\bm{Q}=(0,0,1/2)$ at $\tn=2.8$ K\cite{JPSJ61-1461},
 with the saturation moment of
$m_0 \sim 0.38 \mu_B$\cite{JPSJ61-4667}. 
Note that $ (Q_a,Q_b,Q_c)$ denotes $Q_a a^*+ Q_b b^*+ Q_c c^*$, where $a^*$, $b^*$, and $c^*$
are the reciprocal lattice vectors.  
The change in the ordering vector is observed 
in Ce(Pd$_{1-x}$M$_x$)$_2$Al$_3$,
in which Ag or Cu are substituted for Pd\cite{PRB70-174429, PhB359-287, JPCM18-5715}.
The $(0,0,1/2)$-AF order region  $x<0.05$ is followed by the crossover region $0.05<x<0.1$, 
the ferromagnetic (F) order region $0.1<x<0.4$, another crossover region $0.4<x<0.5$, and 
another AF order (the ordering vector is unreported) region $0.5<x$. 
At $x=0.2$, the reversed anisotropy in the susceptibility of $\chi_a /\chi_c<1$ 
is observed\cite{PRB70-174429, PhB359-287}, indicating that the  CEF ground state 
has changed from the $M=\pm 1/2$ states to other states. 
Although the susceptibility cannot be fit well,
it is natural to assume the  CEF ground state for the doped system as $M=\pm 3/2$, 
because the estimated splitting from the ground state $M=\pm 1/2$  
is  small in the undoped system (CePd$_2$Al$_3$). 
Moreover, the substitution of Ag or Cu gives rise to the electron doping in the conduction band. 
To understand the ordering-vector variation, 
we should consider the realistic conduction-band structures, as well as  the charge distribution anisotropy
of the $f$ states. 
Here we report a theoretical study on the magnetism of Ce(Pd$_{1-x}$M$_x$)$_2$Al$_3$ 
using the anisotropic RKKY interaction which reflects  both the spacial anisotropic distribution of the $f$ states 
and the conduction-band structures.

With lowering temperature ($T$), the $f$ electrons in heavy-fermion compounds 
exhibit the crossover from the localized to itinerant state, 
as observed in the angle-resolved photoemission spectroscopy(ARPES) 
for CeIrIn$_5$\cite{Scie318-1615} and UPd$_2$Al$_3$\cite{NPh3-618}. %\cite{Scie318-1615,NPh3-618}.
For  $T>T^{\ast} \approx 50$~K,
the $f$ electrons are localized at the energy level $E_f$ far below the Fermi level. 
In the region below $T^{\ast}$, the $f$ electrons become gradually hybridized 
with the conduction band, to %
form the renormalized quasi-particles, namely the heavy fermions. 
Even in this low $T$ region, however, the quasi-particles possess
 very large  mass enhancement factor $m^{\ast}/m \sim z^{-1} $, indicating that
the $f$ electrons with the fraction of $1-z \approx 0.9-0.99$
persist as the incoherent, namely localized, part.  
In this paper we discuss the relative stability of the magnetic states with different ordering vectors. 
In this case, the large fraction of the localized $f$ electrons remaining still at 
$T<T^{\ast}$ dominantly contributes to determining the magnetism, 
and the rest, that is, the small fraction of the itinerant $f$ electrons give only a small deviation on it. 
Therefore, 
 the Ruderman-Kittel-Kasuya-Yosida (RKKY) interaction is responsible for
the magnetism of these heavy-fermion compounds, even though the magnetic transition temperature, typically $ 1-10$~K, is
lower than $T^{\ast}$. 
The magnetic order is considered to be influenced  by the orbital anisotropy 
of the $f$ electrons, which is reflected in the susceptibility anisotropy, and also by the conduction-band structures. 

\section{\label{sec1} Formalism}
We obtain the conduction-band structures of CePd$_2$Al$_3$ 
from the band calculation using abinit\cite{abinit1,abinit2} for
the non-$f$ counterpart system, in which La are substituted for Ce. 
In the band calculation, we adopt the Troullier-Martins pseudopotential\cite{tm1},
and the exchange-correlation term is determined according to Perdew and Wang\cite{perdew}. 
The wave functions are expanded by the plane waves up to the cutoff energy of 60 Ry, 
and the 12$^3$ $\bm{k}$-point mesh with the special point technique
by Monkhorst and Pack\cite{special} are adopted.

Using this conduction-band structures, we calculate the RKKY interactions
between $4f$ electrons of the Ce ions. 
In this case, it is necessary to take into account 
the effects of the anisotropic charge distribution of 
the $4f$ electrons on the hybridization, as initiated
by Coqblin and Schrieffer.\cite{Coqblin-Schrieffer} 
Instead of the on-site hybridizations between the conduction and $f$ electrons 
(c-$f$ hybridizations) with the plane-wave conduction states
in Ref. \onlinecite{Coqblin-Schrieffer}, 
we adopt the c-$f$ hybridizations 
derived from the two-center Slater-Koster integrals\cite{slater,takegahara}
between the Ce and Pd (or Al) sites in the following way. 
We take account of all conduction bands (labeled by $\gamma$) 
which cross the Fermi level, 
where only the Al 3$p$, Pd $5p$, and Pd $4d$ electrons are assumed to 
hybridize with the $f$ electrons on the Ce sites.  
For the Ce $4f$ electrons, 
we consider only a single ground-state CEF doublet labeled by 
the $z$ component of $J$, $M=\pm|M|$ ($|M|=1/2$, 3/2, or 5/2). 

Taking the number of unit cells $N$, the RKKY-interaction Hamiltonian is expressed as 
\begin{align}
H_\mathrm{RKKY} &= %
\frac{1}
{{2N}}\sum\limits_{ijMM' } {E_{|M|} 
\left( \mu, {{\bm{R}}_i  - {\bm{R}}_j } \right)f_{iM} ^\dag  f_{iM'} f_{jM'} ^\dag  f_{jM} } 
, \label{eqhami0} \\
%%%%%%%%%%%%%%%%%%%%%%%%%%%%%%%%%%%%%%%%%%
%
E_{\left| M \right|} \left(\mu, {\bm{R}} \right) &= \frac{2}
{{E_f ^2 }}\sum\limits_{{\gamma\bm{kk'}} } {\frac{{f\left( {\varepsilon _{\gamma {\bm{k}}} }
-\mu \right)\left[ {1 - f\left( {\varepsilon _{\gamma {\bm{k'}}} } -\mu \right)} \right]}}
{{\varepsilon _{\gamma {\bm{k}}}  - \varepsilon _{\gamma {\bm{k'}}} }}} 
\nonumber \\
%%%%%%%%%%%%%%%%%%%%%%%%%%%
& 
\times
B_{\gamma {\bm{k}}M }  e^{ - \rmi{\bm{k}} \cdot {\bm{R}}} 
B_{\gamma {\bm{k}'}M }  e^{\rmi{\bm{k'}} \cdot {\bm{R}}} 
\nonumber \\
%%%%%%%%%%%%%%%%%%%%%%%%%%%%%%%%%%
&= \frac{2}
{{E_f ^2 }}\sum\limits_{{\gamma\bm{kk'}} } {\frac{{f\left( {\varepsilon _{\gamma {\bm{k}}} }-
\mu \right)\left[ {1 - f\left( {\varepsilon _{\gamma {\bm{k'}}} } -\mu \right)} \right]}}
{{\varepsilon _{\gamma {\bm{k}}}  - \varepsilon _{\gamma {\bm{k'}}} }}} 
\nonumber \\
%%%%%%%%%%%%%%%%%%%%%%%%%%%
& 
\times
B_{\gamma {\bm{k}}M } \cos({ {\bm{k}} \cdot {\bm{R}}} ) 
B_{\gamma {\bm{k}'}M }  \cos({{\bm{k'}} \cdot {\bm{R}}} )
, \label{eqhami1} 
\end{align}
where $\mu$, $\varepsilon _{\gamma \bm{k}}$, and $f(\varepsilon _{\gamma \bm{k}}-\mu)$
 are the chemical potential,  the energy of the conduction band $\gamma$, and 
the Fermi distribution function, respectively. 
$B_{\gamma {\bm{k}}M} $
consists of the terms derived from all process in which the $4f$ electrons on a Ce sites and 
the conduction electron  are scattered into the $4f$ electrons on the Ce sites  and conduction electrons
with their angular momenta unchanged. 
The conduction electrons with given  $l$ and $m$ are derived from
various crystallographically inequivalent atoms, denoted as Ce, Al1, Al2, Al3, Pd1, and Pd2. 
$B_{\gamma {\bm{k}}M} $ includes two kinds of terms derived 
from two kinds of scattering: 
(1) the scattering in which
the conduction electrons from equivalent atom are concerned;
(2) the scattering in which
the conduction electrons from inequivalent atoms are concerned. 
Here, we denote (1) as the equivalent-atom scattering and (2) as the 
inequivalent-atom scattering. 
To calculate the the inequivalent-atom terms in $B_{\gamma {\bm{k}}M} $, 
we may be required to 
obtain with high accuracy the composition of 
the conduction electron using the decomposition into Wannier functions
on the basis of the Slater-Koster method. 
Such a task is, however, difficult to execute. 
To avoid this difficulty, we replace the sum of such inequivalent-atom terms
in $B_{\gamma {\bm{k}}M} $
by a constant $V_0^2$, 
because the sum of such terms are expected to consist of variously $\bm{k}$-dependent term, 
and hence to have a weak $\bm{k}$ dependence. 
On the other hand, 
for the equivalent-atom scattering, we calculate the corresponding terms in 
$B_{\gamma {\bm{k}}M} $
by the use of the two-center 
Slater-Koster integrals.\cite{slater,takegahara}

Thus, we obtain the expression of $B_{\gamma {\bm{k}}M}$ as
%%%%%%%%%%%%%%%%%%%%%%%%%%%%%%%%%%
\begin{align}
B_{\gamma {\bm{k}}M}  = 4 \sum\limits_{X\sigma nl m}^{} 
{{\rho _{\gamma \bm{k} Xnlm} }\frac{{7 - 4M\sigma }}
{{14}}\left| {V_{\gamma {\bm{k}}M - \sigma Xnlm} } \right|^2 } + V_0^2
\label{eqb2} 
,
\end{align}
%%%%%%%%%%%%%%%%%%%%%%%%%%%%%%%%%%
where $X$ is the label of crystallographically inequivalent atoms, 
$\sigma$ is the spin of the $f$ electron, and
$n$, $l$, $m$ are the principal, azimuthal, and magnetic quantum numbers 
of the conduction electrons, respectively. 
$V_{\gamma {\bm{k}}M - \sigma Xlm} $ is the hybridization matrix element obtained from the 
Slater-Koster integrals $I_{M - \sigma Xnlm} \left( {{\bm{r}}_j } \right)$;
%%
%%%%%%%%%%%%%%%%%%%%%%%%%%%%%%%%%%
\begin{align}
V_{\gamma {\bm{k}}M - \sigma Xnlm}  = \sum\limits_{j \in X}^{} {I_{M - \sigma Xnlm} 
\left( {{\bm{r}}_j } \right)e^{-{\text{i}}{\bm{k}} \cdot {\bm{r}}_j } } 
\label{eqhybxx} ,
\end{align}
%%%%%%%%%%%%%%%%%%%%%%%%%%%%%%%%%%
where $j$ denotes  all sites of $X$ atoms, which are distant by ${\bm{r}}_j $
from the relevant Ce site. 
The coefficient
${\rho _{\gamma \bm{k} Xnlm} }$ 
cannot be obtained directly from our band calculation. 
What we can obtain is only the partial density of states (DOS) of the conduction electrons
derived from given kind of atom (Ce, Pd, or Al), with given $n$ and $l$. 
Therefore, we assume that
$\rho_{\gamma \bm{k} \mathrm{Al}i\,3p \,m}=P_{\mathrm{Al}3p}/9$, 
where 
$P_{\mathrm{Al}3p}$ is the ratio of the partial DOS of the Al $3p$ electrons to the total conduction-electron 
DOS at the Fermi level, 
and the denominator 9 is the product of 3 (Al1-3) and 3 ($m=1,0,-1$). 
Similarly, we assume that
$\rho_{\gamma \bm{k} \mathrm{Pd}i\,5p \,m}=P_{\mathrm{Pd}5p}/6$ 
and
$\rho_{\gamma \bm{k} \mathrm{Pd}i\,4d \,m}=P_{\mathrm{Pd}4d}/10$. 
The value obtained from our band calculation is
$P_{\mathrm{Al}3p}=0.27$, 
$P_{\mathrm{Pd}5p}=0.20$, and
$P_{\mathrm{Pd}4d}=0.15$. 

We discuss, as an example, how to obtain 
$\left| {V_{\gamma {\bm{k}}M - \sigma \mathrm{Al1}3p\,m} } \right|^2$. 
In this case, neighboring
Al1 atoms are distant from the relevant Ce site by $\bm{r}_1= \ffrac{\bm{a}}{2}+\ffrac{\bm{c}}{2}$, 
$\bm{r}_2= -\ffrac{\bm{a}}{2}+\ffrac{\bm{c}}{2}$, 
$\bm{r}_3 \equiv -\bm{r}_1= -\ffrac{\bm{a}}{2}-\ffrac{\bm{c}}{2}$, 
and $\bm{r}_4 \equiv -\bm{r}_2= \ffrac{\bm{a}}{2}-\ffrac{\bm{c}}{2}$.  
In the case that $M-\sigma-m$ is even, 
the Slater-Koster integrals  exhibit the same value
 for the four distance $\bm{r}_\text{1-4}$, giving rise to  
%%%%%%%%%%%%%%%%%%%%%%%%%%%%%%%%%%
\begin{align}
|{V_{\gamma {\bm{k}}M\sigma \mathrm{Al1} 3p\,m} }|^2&
=16 {\left| {I_{M-\sigma\mathrm{Al}1\,3p\,m} }{(\bm{r}_1)}  \right|^2 } 
\cos^2\left({\frac{\bm{a}}{2}\cdot\bm{k}}\right)\cos^2\left({\frac{\bm{c}}{2}\cdot\bm{k}} \right)
\label{eqv21a} .
\end{align}
%%%%%%%%%%%%%%%%%%%%%%%%%%%%%%%%%% 
%%
For odd $M-\sigma-m$, 
the Slater-Koster integrals  exhibit the same value
for  $\bm{r}_{1,3}$ and the value multiplied by $(-1)$ for $\bm{r}_{2,4}$, giving rise to 
%%%%%%%%%%%%%%%%%%%%%%%%%%%%%%%%%%
\begin{align}
|{V_{\gamma {\bm{k}}M\sigma \mathrm{Al1} 3p\,m} }|^2&
=16 {\left| {I_{M-\sigma\mathrm{Al}1\,3p\,m} } {(\bm{r}_1)} \right|^2 } 
\sin^2\left({\frac{\bm{a}}{2}\cdot\bm{k}}\right)\sin^2\left({\frac{\bm{c}}{2}\cdot\bm{k}}\right)
\label{eqv21b} .
\end{align}
%%%%%%%%%%%%%%%%%%%%%%%%%%%%%%%%%% 
%%
Similarly, 
$
|{V_{\gamma {\bm{k}}M\sigma \mathrm{Al2\,(3)} } }|^2$
is obtained by replacing $\bm{a}$ by $\bm{b}$  ($\bm{a}+\bm{b}$) in
Eqs.~(\ref{eqv21a})-(\ref{eqv21b}), 
noting that $|I_{M - \sigma \mathrm{Al}i\,lm} \left( {{\bm{r}}_j } \right)|^2 $ 
is independent of $i$, and of $j$, which holds for Pd as well. 
Thus, $
\sum\limits_i^{} {\left| {V_{\gamma {\bm{k}}M - \sigma {\text{Al}}i\,3p\,m} } \right|^2 } 
$ is large at
$\Gamma$ (L, and H) for even (odd) $M-\sigma-m$. 
We can give similar discussion on Pd1 (distant from Ce by 
$\bm{r}'_1=2\bm{a}/3+\bm{b}/3$, $\bm{r}'_2=-\bm{a}/3+\bm{b}/3$, 
and $\bm{r}'_3=-\bm{a}/3-2\bm{b}/3$) and Pd2 
(distant by $-\bm{r}'_1$, $-\bm{r}'_2$, and $-\bm{r}'_3$), and
obtain large 
$
\sum\limits_i^{} {\left| {V_{\gamma {\bm{k}}M - \sigma {\text{Pd}}i\,5p\,m} } \right|^2 } 
$
and $
\sum\limits_i^{} {\left| {V_{\gamma {\bm{k}}M - \sigma {\text{Pd}}i\,4d\,m} } \right|^2 } 
$
at the U and P line for $M-\sigma-m \neq 0$ mod 3 and 
at the $\Delta$ line for $M-\sigma-m = 0$ mod 3. 

In this way,
$B_{\gamma {\bm{k}}M} $
is determined by $\left| {I_{M-\sigma\mathrm{Al}1\,3p\,m} } \right|^2$, 
$\left| {I_{M-\sigma\mathrm{Pd}1\,5p\,m} } \right|^2$, 
and $\left| {I_{M-\sigma\mathrm{Pd}1\,4d\,m} } \right|^2$, 
dependent on
the $z$-direction two-center Slater-Koster integrals \cite{slater,takegahara}
$(fp\sigma)_\mathrm{Al}$, $(fp\pi)_\mathrm{Al}$, 
$(fp\sigma)_\mathrm{Pd}$, $(fp\pi)_\mathrm{Pd}$, etc. 
We adopt $(fp\sigma)_\mathrm{Al}$, $(fp\sigma)_\mathrm{Pd}$, 
$(fd\sigma)_\mathrm{Pd}$, and $V_0$ as the independent parameters, 
by assuming for simplicity that $(fl\pi)_X/(fl\sigma )_X=-0.5$
and $(fd\delta)_\mathrm{Pd}=0$.

We further note that the charge distribution of the $f$-electron states affects the anisotropy
of the c-$f$ hybridizations, 
through the factor $(7-4M\sigma))/14$ in Eq.~(\ref{eqb2}) and
the two-center Slater-Koster integrals. 
For example, the $f$ states with $M=\pm 1/2$  have a small charge distribution in 
the basal plane, and hence a weak hybridization with the $5p$ and 4$d$ states at Pd sites 
located in the basal plane, 
but have a strong hybridization with $3p$ states at Al sites. 
The resulting $\bm{k}$ dependence of
$B_{\gamma {\bm{k}}1/2} $ shows that it
has large values near the  $\Gamma$ point. 
Similarly, 
$B_{\gamma {\bm{k}}3/2} $ has large values near  the $\Gamma$, $L$, and $H$ points. 
On the other hand, the
$f$ states with $M=\pm 5/2$  strongly hybridize with the $5p$ and 4$d$ states at Pd sites, 
and consequently we obtain
$B_{\gamma {\bm{k}}5/2} $ having large values near the U and P lines.

Since we are interested in the sufficiently low temperature region compared with the conduction bandwidth, 
${f\left( {\varepsilon _{\gamma \bm{k}}  }-\mu \right)}$ can be approximated 
by the step function $\theta( \mu -\varepsilon _{\gamma \bm{k}}  )$, 
where $E_{\left| M \right|} \left( \mu,{\bm{R}} \right) $ is rewritten as
\begin{align}
E_{\left| M \right|} \left( \mu, {\bm{R}}   \right) &= \frac{2}
{{E_f ^2 }}\sum\limits_\gamma ^{} {\int_{\varepsilon _{\gamma \min } }^\mu
  {d\varepsilon } \int_\mu ^{\varepsilon _{\gamma \max } } {d\varepsilon '} 
\frac{{g_{\gamma \left| M \right|} \left( \varepsilon ,{\bm{R}} \right)
g_{\gamma \left| M \right|} \left( {\varepsilon ',{\bm{R}}} \right)}}
{{\varepsilon  - \varepsilon '}}} 
, \label{eqrk1} \\
%%%%%%%%%%%%%%%%%%%%%%%%%%%%%%%%%%%%%%%%%
g_{\gamma |M|} \left( \varepsilon ,{\bm{R}} \right) &= 
\int_{\varepsilon _{\gamma {\bm{k}}}  = \varepsilon }^{} {\frac{{dS}}
{{\left| {\partial \varepsilon _{\bm{k}} /\partial {\bm{k}}} \right|}}
B_{\gamma {\bm{k}}|M|} \cos \left( {{\bm{k}} \cdot  {\bm{R}}} \right)} 
\nonumber \\
%%%%%
&= 
\frac{1}{N_{D_{6h}}}
\int_{\varepsilon _{\gamma {\bm{k}}}  = \varepsilon }^{} {\frac{{dS}}
{{\left| {\partial \varepsilon _{\bm{k}} /\partial {\bm{k}}} \right|}}B_{\gamma {\bm{k}}|M| } 
\sum\limits_{\bm{k}_{\mathrm{eq}}} \cos \left[{{\bm{k}_{\mathrm{eq}}}} \cdot  {\bm{R}} \right]
} 
, \label{eqrk2} 
\end{align}
where $
\int_{\varepsilon _{\gamma \bm{k}}  = \varepsilon }^{} {dS} 
$
denotes the integration over the iso-energy surface 
${\varepsilon _{\gamma \bm{k}}  = \varepsilon }$ in the
first Brillouin zone (FBZ).  
$\varepsilon _{\gamma \max }$ and $\varepsilon _{\gamma \min }$ are  
the maximum and minimum energies 
of the conduction band $\gamma$, respectively. 
$\bm{k}_{\mathrm{eq}}$ are the points in the FBZ, including $\bm{k}$  itself, 
to which $\bm{k}$ is translated by the symmetry operations
belonging to the ${D_{6h}}$ point group, of which
the number of the elements 
${N_{D_{6h}}}=24$. 

Using the sine transformation 
$g_{\gamma |M|} \left( \varepsilon, {\bm{R}} \right) = \sum\limits_m^{} {a_{m|M|} ({\bm{R}})}     
\sin \left( {m\pi \left( {\varepsilon  - \varepsilon _{\gamma \min } } \right)/W_\gamma  } \right)
$, 
where $
W_\gamma   = {\varepsilon _{\gamma \max }  - \varepsilon _{\gamma \min } }  $ 
is the width of the conduction band $\gamma$, 
$E_{\left| M \right|} \left( \mu,{\bm{R}}  \right) $ is expressed as 
\begin{align}
E_{\left| M \right|} \left( \mu, {\bm{R}} \right) &=\sum\limits_\gamma ^{}
\frac{ W_\gamma   }{\pi E_f^2}
\sum\limits_{m,n}^{} {a_{m|M|} ({\bm{R}}) a_{n|M|} ({\bm{R}}) }
\nonumber \\
%%%%%%%%%%%%%%%%%%%%%%%%%%%
& \hspace{0.2cm}
\times
{\left[ {I_{m, - n} \left( {\frac{{\mu  - \varepsilon _{\gamma \min } }}
{{W_\gamma  }}\pi } \right) - I_{m,n} \left( {\frac{{\mu  - \varepsilon _{\gamma \min } }}
{{W_\gamma  }}\pi } \right) } \right]} 
,  \label{eqrk3} \\
%%%%%%%%%%%%%%%%%%%%%%%%%%%
I_{m,n} \left( \nu  \right) &= \mathrm{Re} \int_0^\nu  d u\int_\nu ^\pi  {du'} 
\frac{{\exp \left( {{\mathrm{i}}mu} \right)\exp \left( {{\mathrm{i}}nu'} \right)}}
{{u - u'}}
. \label{eqrk4} 
%%%%%%%%%%%%%%%%%%%%%%%%%%%%%%%%%%%%%%%
\end{align}
$I_{m,n} \left( \nu  \right) $
is analytically calculated as 
\begin{align}
I_{m,n \ne  - m} \left( \nu  \right) &= \frac{{\sin \left( {\left( {m + n} \right)\nu } \right)}}
{{m + n}}\left[ {{\mathrm{Ci}}\left( {\left| m \right|\nu } \right) - \log \left| m \right| 
- {\mathrm{Ci}}\left( {\left| n \right|\left( {\pi  - \nu } \right)} \right) + \log \left| n \right|} \right]
\nonumber \\
%%%%%%%%%%%%%%%%%%%%%%%%%%%%%%%%%%%%%%%%
 &%
- \frac{{\cos \left( {\left( {m + n} \right)\nu } \right)}}
{{m + n}}\left[ {{\mathrm{Si}}\left( {m\nu } \right) 
+ {\mathrm{Si}}\left( {n\left( {\pi  - \nu } \right)} \right)} \right]
\nonumber \\
%%%%%%%%%%%%%%%%%%%%%%%%%%%%%%%%%%%%%%%%
 &%\hspace{1.0cm} 
- \frac{{{\mathrm{Si}}\left( {n\nu } \right) - {\mathrm{Si}}\left( {n\pi } \right) 
+ \left( { - 1} \right)^{m + n} \left[ {{\mathrm{Si}}\left( {m\left( {\pi  - \nu } \right)} \right) 
- {\mathrm{Si}}\left( {m\pi } \right)} \right]}}
{{m + n}}
, \label{eqimn1} \\
%%%%%%%%%%%%%%%%%%%%%%%%%%%%%%%%%%%%%%%%%
I_{m, - m} \left( \nu  \right) &= \frac{{\left[ {\left( { - 1} \right)^m  - 1} \right]
\sin \left( {m\nu } \right)}}
{m} 
\nonumber \\
%%%%%%%%%%%%%%%%%%%%%%%%%%%%%%%%%%%%%%%%
 &%\hspace{1.0cm} 
+ \nu {\mathrm{Ci}}\left( {\left| m \right|\nu } \right) + 
\left( {\pi  - \nu } \right){\mathrm{Ci}}\left( {\left| m \right|\left( {\pi  - \nu } \right)} \right) 
- \pi {\mathrm{Ci}}\left( {\left| m \right|\pi } \right)
,  \label{eqrimn2} 
%%%%%%%%%%%%%%%%%%%%%%%%%%%%%%%%%%%%%%%
\end{align}
where ${\mathrm{Ci}}(x)$ and ${\mathrm{Si}}(x)$ are the cosine and sine integral functions. 
Thus, calculating numerically $g_{\gamma |M|} \left( \varepsilon  \right) $, 
which reflects the Fermi surface topology and the hybridization term anisotropy, 
we obtain the RKKY interaction. 
In the mean-field approximation (MFA), 
Eq.~(\ref{eqhami0}) 
is rewritten by the use of the Fourier transformations $
f_{iM} ^\dag   = \sum\limits_{\bm{k}} {f_{{\bm{k}}M} ^\dag  
\expu{\rmi{\bm{k}} \cdot {\bm{R}}_i } } 
$ and
${
f_{iM}    = \sum\limits_{\bm{k}} {f_{{\bm{k}}M}   \expu{-\rmi{\bm{k}} \cdot {\bm{R}}_i } } 
}$
as 
%%
%%%%%%%%%%%%%
\begin{align}
H_\mathrm{RKKY}  &=
% =\sum\limits_{{\bm{k}}M} {\varepsilon ^f_{|M|} f_{{\bm{k}}M} ^\dag f_{{\bm{k}}M} } +
 \frac{1}{2N}
\sum\limits_{{\bm{Qk}}_1 {\bm{k}}_2 MM'} {K_{|M|{\bm{Q}}}
f_{{\bm{k}}_1  + {\bm{Q}}M} ^\dag  f_{{\bm{k}}_1 M'} f_{{\bm{k}}_2 M'} ^\dag  
f_{{\bm{k}}_2  + {\bm{Q}}M} }
%\hc
, \label{eq6} \\
%%%%%%%%%%%%%%%%%
 K_{|M|{\bm{Q}}} &= \frac{1}
{N}\sum\limits_{ij} {E_{|M|} ( {\mu,{\bm{R}}_i  - {\bm{R}}_j } ) \exp[{{\mathrm{i}}
{\bm{Q}}  \cdot ( {{\bm{R}}_i  - {\bm{R}}_j } )  } } ]
, \label{eq61} 
\end{align}
%%%%%%%%%%%%%%%. 
%%
and
the magnetic order parameter $\Delta_{ {|M|}\bm{Q}}$ for the ordering vector $\bm{Q}$ 
satisfies the self-consistent equation 
\begin{align}
\Delta _{{|M|}{\bm{Q}}}  =  - \frac{{K_{{|M|}{\bm{Q}}} }}
{2}\tanh \frac{{\Delta _{{|M|}{\bm{Q}}} }}
{{2T}}.  \label{eqgap} 
\end{align}
Thus, the realizing ordering vector is given by  $\bm{Q}_{\rm{min}}$, 
at which ${K_{{|M|}{\bm{Q}}} }$ takes the minimum $K_{{|M|}{\bm{Q}}_{\min } } $, 
giving rise to the transition temperature $\tn=-
{K_{{|M|}{\bm{Q}}_{\rm{min}}} }/4$.  
Generally, if
the RKKY interaction $E_{\left| M \right|} \left( \mu,{\bm{a}} \right) $ between
 the in-plane neighboring sites 
is positive, 
the magnetic order with nonzero in-plane component  of $\bm{Q}$ (namely, $Q_a$ and $Q_b$) 
tends to be stable. 
On the other hand, if $E_{\left| M \right|} \left( \mu,{\bm{c}} \right) $ between 
the $c$-direction neighboring sites 
is positive, 
the order with nonzero $Q_c$ tends to be stable.

Next, we discuss the relation between the iso-energy integrations 
$g_{\gamma |M|} \left( \varepsilon, \bm{R}  \right) $ and the ordering vector. 
Since
\begin{align}
{\int_{\varepsilon _{{\varepsilon_0}-D} }^{\varepsilon_0}  {d\varepsilon } 
\int_{\varepsilon_0} ^{{\varepsilon_0}+D} {d\varepsilon '} \frac{{g_{\gamma |M|}
\left( \varepsilon  \right)g_{\gamma |M|}\left( {\varepsilon '} \right)}}
{{\varepsilon  - \varepsilon '}}} 
\nonumber \\
%%%%%%%%%%%%%%%%%%%%%%%%%%%%%%%%
  &\hspace{-5.0cm}
\simeq \int_{ - D}^0 {ds} \int_0^D {ds'} \frac{{\left( {g_{\gamma |M|}\left( {\varepsilon_0}  \right) 
+ g'_{\gamma |M|}\left( {\varepsilon_0}  \right)s} \right)\left( {g_{\gamma |M|}
\left( {\varepsilon_0}  \right) + g'_{\gamma |M|}\left( {\varepsilon_0}  \right)s'} \right)}}
{{s - s'}}
\nonumber \\
%%%%%%%%%%%%%%%%%%%%%%%%%%%%%%%%
 &\hspace{-5.0cm}
=  - 2\log 2\left[ {g_{\gamma |M|}\left( {\varepsilon_0}  \right)} \right]^2 D + \frac{{(2 - 2\log 2)}}
{3}\left[ {g'_{\gamma |M|}\left( {\varepsilon_0}  \right)} \right]^2 D^3 
, \label{eqkinji} 
\end{align}
%%%%%%%%%%%%%%%%%%%%%
$E_{\left| M \right|} \left({\varepsilon_0}, {\bm{R}}  \right) $ in Eq.~(\ref{eqrk1}) 
is expected to increase with
decreasing $g_{\gamma |M|} \left( {\varepsilon_0} ,{\bm{R}} \right) $ 
and with increasing $g'_{\gamma |M|} \left( {\varepsilon_0} ,{\bm{R}} \right) $, 
and therefore to have a local maximum 
with respect to $\varepsilon$, in the case that 
$g_{\gamma |M|} \left( \varepsilon ,{\bm{R}} \right) $ changes its sign
near $\varepsilon={\varepsilon_0}$. 

The sign of $g_{\gamma |M|} \left( \varepsilon ,{\bm{R}} \right) $ in Eq.~(\ref{eqrk2}) 
is determined mainly by
the topology of the iso-energy surface ${\varepsilon _{\gamma \bm{k}}  = \varepsilon }$. 
For $\bm{R}=\bm{a}$, where $\bm{a}$ is the in-plane lattice vector, 
 $\sum\limits_{\bm{k}_{\mathrm{eq}}} \cos \left[{{\bm{k}_{\mathrm{eq}}}} 
\cdot  {\bm{a}} \right] >0$ for $\bm{k}$ near the $\Delta$ line (the central axis of the FBZ), 
and   $\sum\limits_{\bm{k}_{\mathrm{eq}}} \cos \left[{{\bm{k}_{\mathrm{eq}}}} 
\cdot  {\bm{a}} \right] <0$ for $\bm{k}$ near the outer edge of the FBZ. 
Therefore, if the iso-energy surface moves from near the $\Delta$ line toward the outer edge 
with increasing or decreasing $\varepsilon$, 
$g_{\gamma |M|} \left( \varepsilon ,{\bm{a}} \right) $ changes its sign, 
and $E_{\left| M \right|} \left( {\varepsilon},{\bm{a}} \right) $ is 
expected to have a local maximum. 
If the local maximum $E_{\left| M \right|} \left( {\varepsilon_\mathrm{max}},{\bm{a}} \right) $
 is positive, 
the magnetic order with finite in-plane component of $\bm{Q}$ 
is expected to be stable near $\mu={\varepsilon_\mathrm{max}}$. 
Similarly, $\sum\limits_{\bm{k}_{\mathrm{eq}}} \cos \left[{{\bm{k}_{\mathrm{eq}}}} 
\cdot  {\bm{c}} \right] $ is positive 
near the plane with $k_c=0$ including $\Gamma$ point (the central plane in the FBZ), 
and is negative near the plane $k_c=\pm 1/2$ including the A point
 (the top or bottom planes in the FBZ). 
Therefore, if the iso-energy surface moves from near the central plane toward the top and bottom planes 
with varying  $\varepsilon$, 
$g_{\gamma |M|} \left( \varepsilon ,{\bm{c}} \right) $ change its sign, 
accompanied by a local maximum of $E_{\left| M \right|} \left( {\varepsilon},{\bm{c}} \right) $. 
If the local maximum $E_{\left| M \right|} \left( {\varepsilon_\mathrm{max}},
{\bm{c} } \right) $ 
is positive, 
the magnetic order with finite ${Q_c}$ is expected to be stable. 
In the case that $\varepsilon$ is located near the top or bottom of the band, 
$g_{\gamma |M|} \left( {\varepsilon_0} ,{\bm{R}} \right) $ is large
and $g'_{\gamma |M|} \left( {\varepsilon_0} ,{\bm{R}} \right) $ is small, 
giving rise to the negative
$E_{\left| M \right|} \left( {\varepsilon},{\bm{R}} \right) $ 
which makes the F order stable.  
It should be noted that these relations between the Fermi surface topology and the ordering vector
is quite similar to the nesting effect in the Hubbard model.

\section{\label{sec3} Results and Discussion}

\begin{figure}
\includegraphics[width=8cm]{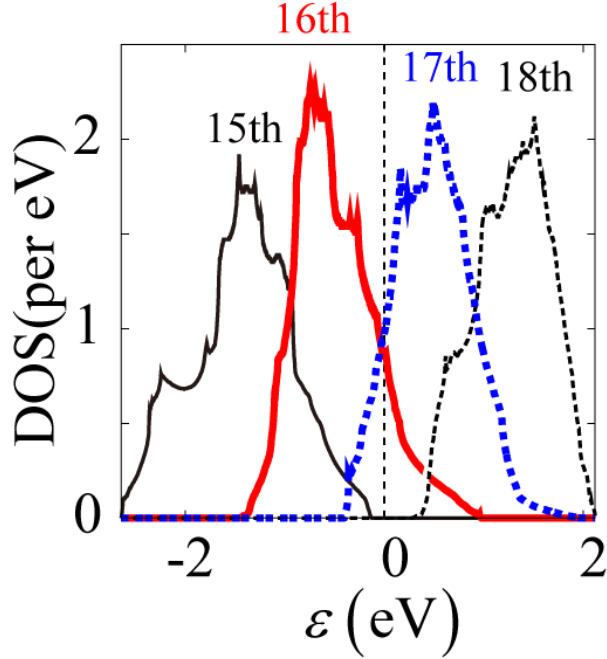}% Here is how to import EPS art
\caption{\label{figdos} (Color online)
The density of states (DOS) versus $\varepsilon$. 
The thin solid, thick solid, thick dashed, and thin dashed lines denote the DOS of the
15th, 16th, 17th, and 18th bands, respectively. 
$\varepsilon=0$ corresponds to the Fermi level for CePd$_2$Al$_3$. 
}
\end{figure}

\begin{figure}
\includegraphics[width=14.5cm]{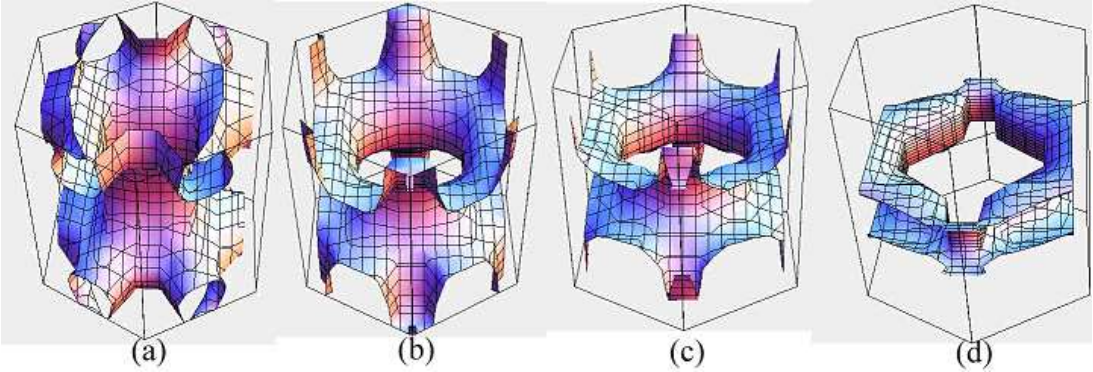}% Here is how to import EPS art
\caption{\label{figiso16} (Color online)
The iso-energy surfaces of the 16th band for (a): $\varepsilon=-0.6$ eV;  (b): $\varepsilon=-0.4$ eV; (c): 
$-0.2$ eV; (d): $0.0$ eV (calculated Fermi surface). 
The $\Gamma$ point is located at the center. 
}
\end{figure}

\begin{figure}
\includegraphics[width=14.5cm]{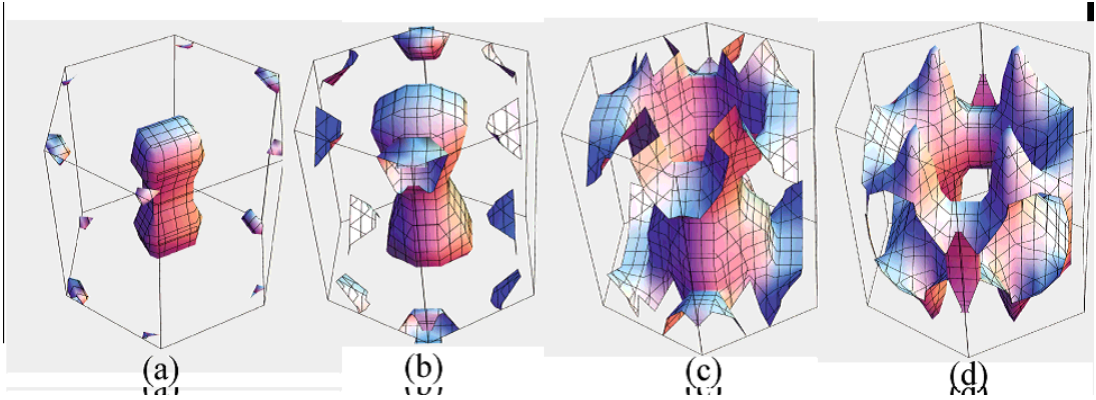} % Here is how to import EPS art
\caption{\label{figiso17} 
(Color online)
The iso-energy surfaces of the 17th band for (a): $\varepsilon=0.0$ eV (calculated Fermi surface); (b): 
$0.2$ eV; (c): $0.4$ eV; (d): 0.6 eV. 
}
\end{figure}

\begin{figure}
\includegraphics[width=10.5cm]{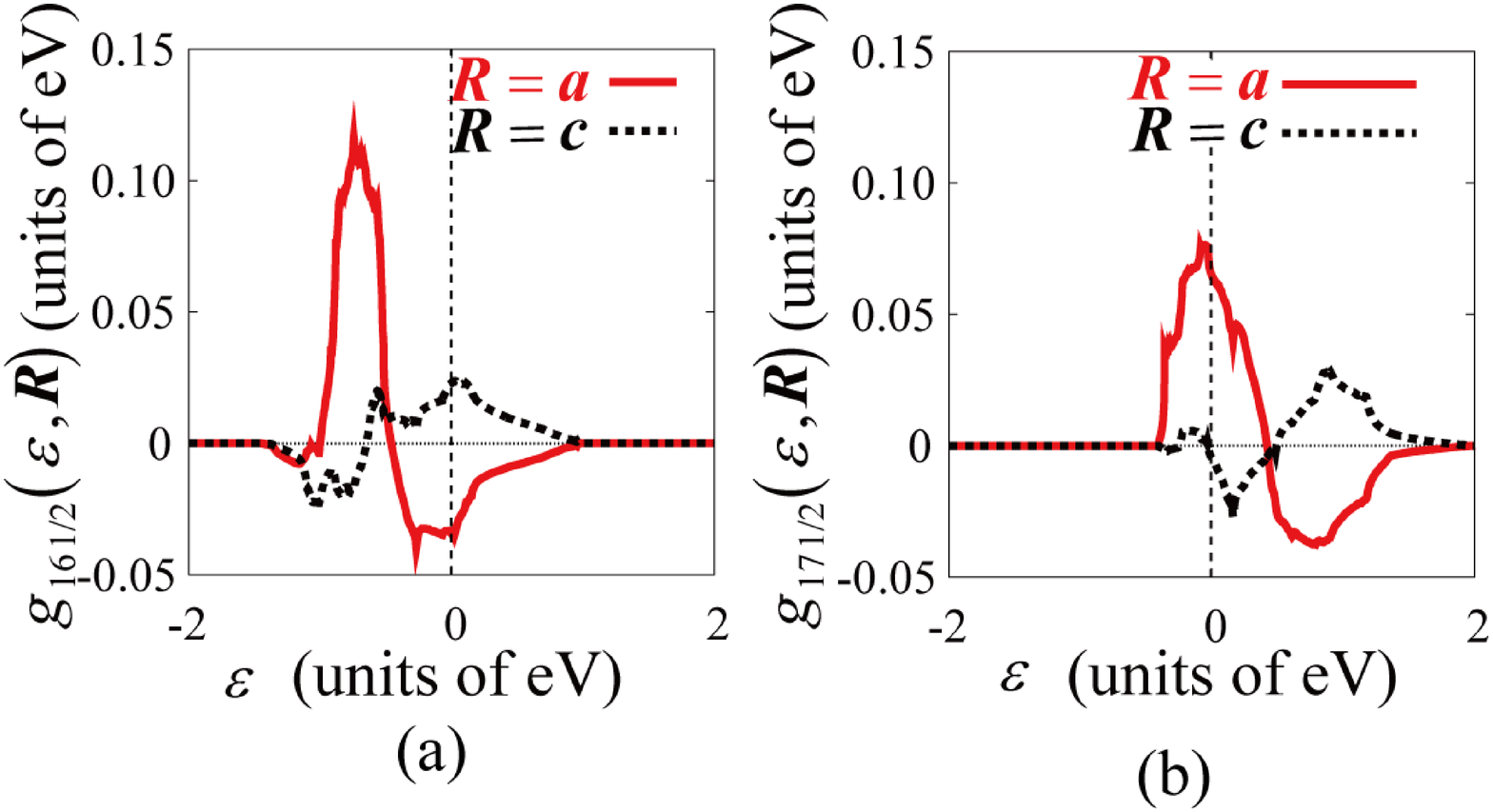} % Here is how to import EPS art
\caption{\label{figofg} 
(Color online)
(a): $g_{16 \,1/2} \left( \varepsilon ,{\bm{R}} \right) $ versus $\varepsilon$;  
(b): $g_{17 \,1/2} \left( \varepsilon ,{\bm{R}} \right) $ versus $\varepsilon$. 
The solid and the dashed lines denote the $\bm{R}=\bm{a}$ 
and $\bm{R}=\bm{c}$, respectively. 
}
\end{figure}

We show the calculation result for 
$(fp\sigma)_\mathrm{Al}=0.21$ eV, $(fp\sigma)_\mathrm{Pd}=0.21$ eV, 
$(fd\sigma)_\mathrm{Pd}=0.14$ eV, $V_0=0.17$ eV, 
and the depth of the bare $f$ level  $E_f=-2.5$ eV.
Among the conduction bands which we obtained, the 16th and 17th bands possess the Fermi surfaces. 
The density of states (DOS) of the 15-18th bands, which are considered in the calculation of
the RKKY interaction, are plotted 
in Fig.~\ref{figdos}. 
It follows that 
CePd$_2$Al$_3$ (the band structures with $\mu=0$) exists in the crossover 
 from the 16th-band dominant region to
the 17th-band dominant one. 
Several of  the iso-energy surfaces for the 16th and 17th bands are plotted in
Figs.~\ref{figiso16}(a)-(d) and \ref{figiso17}(a)-(d), respectively. 
From $\varepsilon=-0.6$ eV to $0.0$ eV, 
the iso-energy surface of the 16th band approaches  the central $a^*b^*$-plane ($\bm{k}=(k_a,k_b,0)$), 
giving rise to the change in the sign of $g_{16 \, 1/2} \left( \varepsilon ,{\bm{c}} \right) $ 
 from negative to positive at 
$\varepsilon \approx  -0.5$ eV as plotted in Fig.~\ref{figofg} (a). 
On the other hand, 
the iso-energy surface of the 17th band moves from the vicinity of the $\Delta$ line 
(the central axis with $\bm{k}=(0,0,k_c)$) to outside 
from $\varepsilon=0.0$ eV to $0.6$ eV, 
and simultaneously approaches  the central $a^*b^*$-plane ($\bm{k}=(k_a,k_b,0)$) at
0.4 eV $<\varepsilon$,
giving rise to the change in the sign of $g_{17 \,1/2} \left( \varepsilon ,{\bm{a}} \right) $  
from positive to negative at 
$\varepsilon  \approx 0.4$ eV
and that of $g_{17 \,1/2} \left( \varepsilon ,{\bm{c}} \right) $
 from negative to positive at 
$\varepsilon \approx  0.5$ eV, as plotted in Figs.~\ref{figofg} (b). 
Because
$g_{\gamma 3/2} \left( \varepsilon ,{\bm{R}} \right) $  is
not largely different from $g_{\gamma 1/2} \left( \varepsilon ,{\bm{R}} \right) $, 
it is not plotted here.

\begin{figure}
\includegraphics[width=10.5cm]{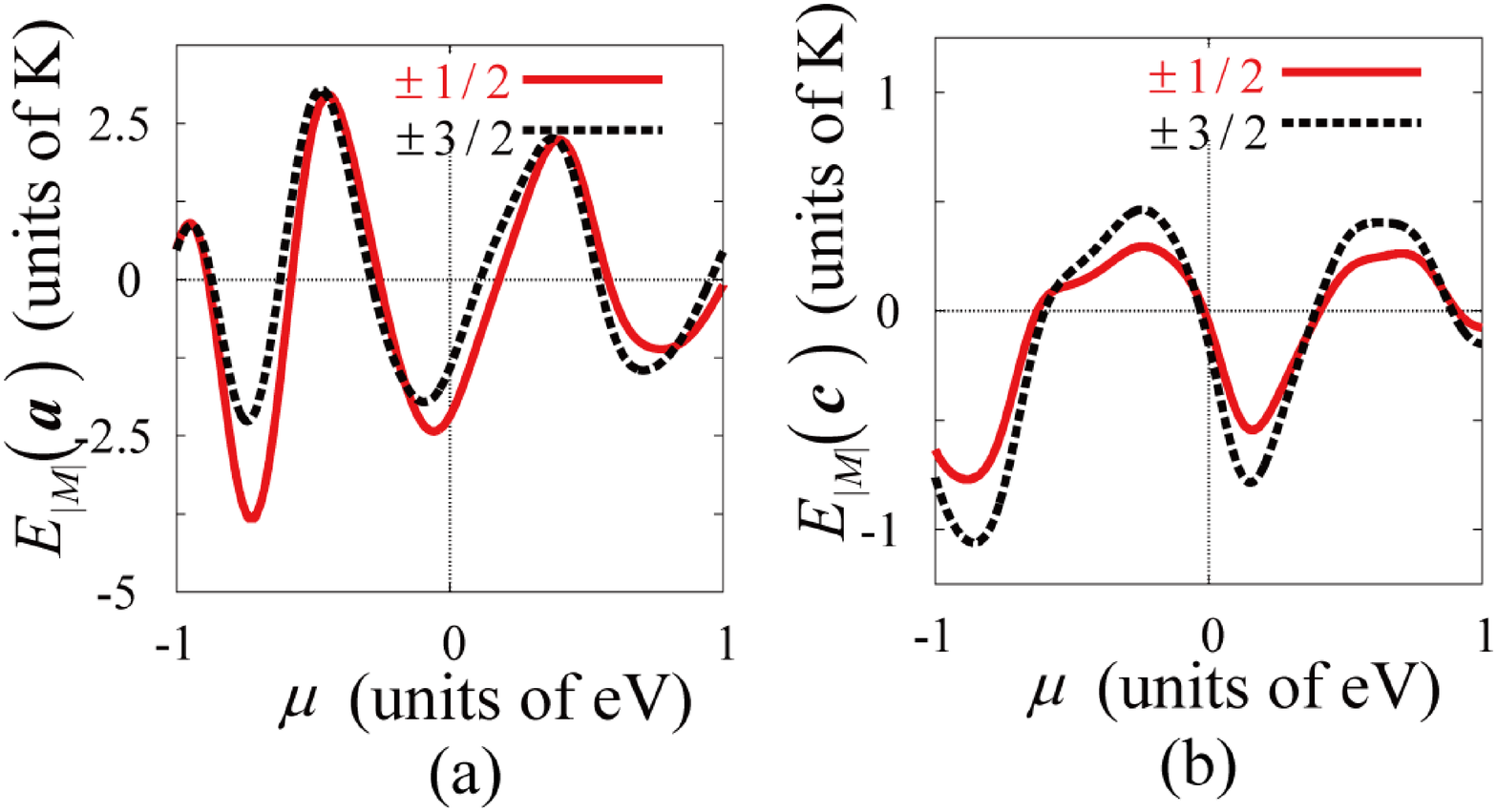} % Here is how to import EPS art
\caption{\label{figpeak} 
(Color online)
(a): $E_{\left| M \right|} \left( {\varepsilon},{\bm{a}} \right) $ versus $\mu$;
(b): $E_{\left| M \right|} \left( {\varepsilon},{\bm{c}} \right) $ versus $\mu$. 
The solid and the dashed lines denote $E_{\left| M \right|}$ for  the $M=\pm 1/2$ and $\pm 3/2$ states, respectively. 
}
\end{figure}

\begin{figure}
\includegraphics[width=10.5cm]{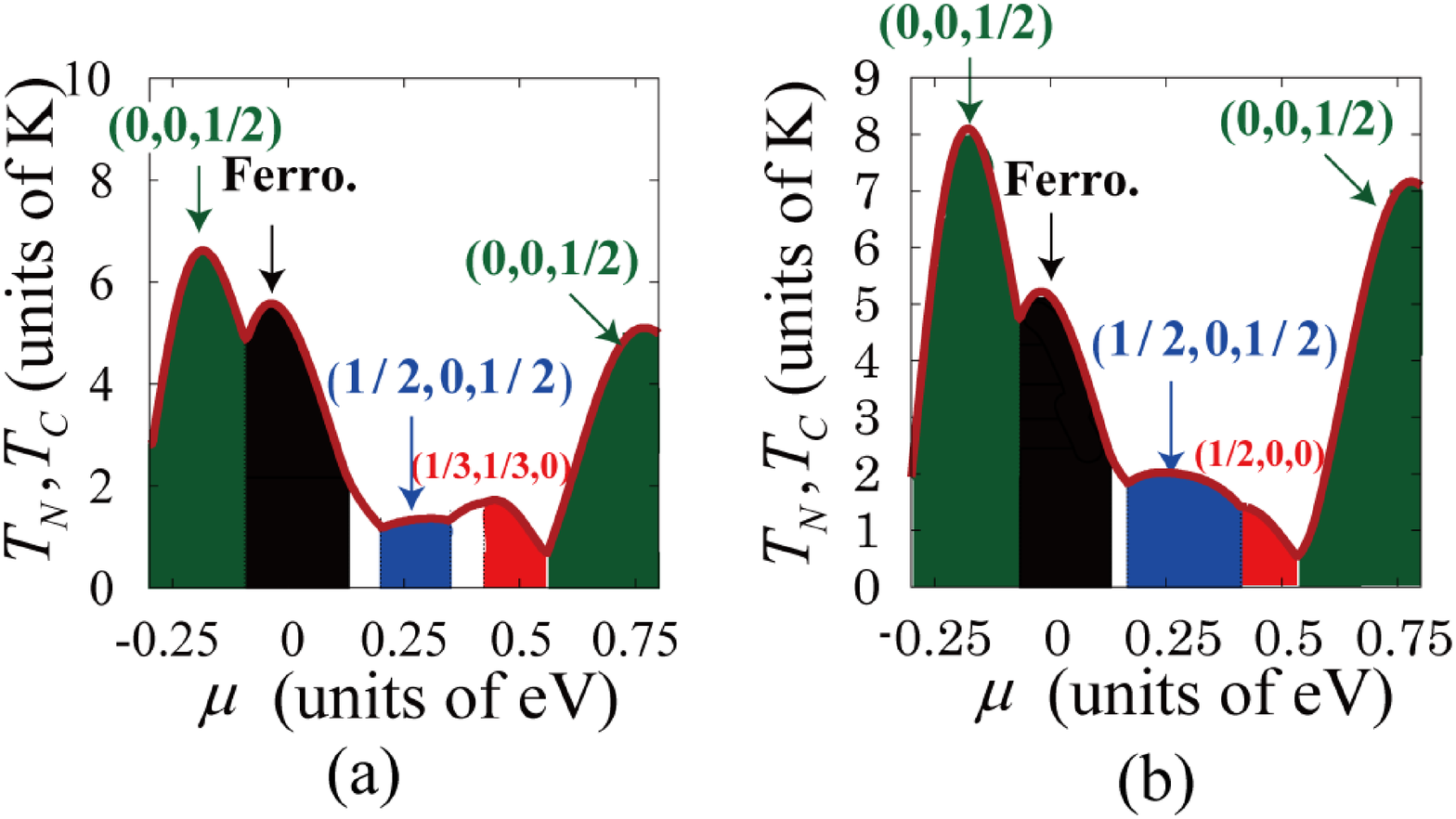} % Here is how to import EPS art
\caption{\label{figphase} 
(Color online)
$\mu$-$T$ phase diagram for 
(a): $M=\pm 1/2$; 
(b): $M=\pm 3/2$. 
}
\end{figure}

These behaviors of $g_{\gamma|M|} \left( \varepsilon ,{\bm{R}} \right) $
give rise to the local maximum of 
$E_{\left| M \right|} \left( {\varepsilon},{\bm{c}} \right) $ near $-0.3$ eV and near  
0.6 eV, and that
 of $E_{\left| M \right|} \left( {\varepsilon},{\bm{a}} \right) $ near 0.3 eV, 
as plotted 
in Figs.~\ref{figpeak} (a) and (b), respectively. 
Thus,  the $(0,0,1/2)$-AF and $(1/2,0,1/2)$-AF states become stable near 
$\mu = -0.2$ eV and near 0.3 eV, 
as plotted in Figs.~\ref{figphase} (a) and (b). 
In the intermediate region $\mu \sim 0$, 
where $\mu$ is located at the crossover  from the 16th dominant region to 17th band dominant one, 
the F order is stable. 

%%%%%%%%%%%%%%%%%%%%%
%%%%%%%%%%%%%%%%%%%%%
%%%%%%%%%%%%%%%%%%%%%
%%%%%%%%%%%%%%%%%%%%%
%%%%%%%%%%%%%%%%%%%%%
%%%%%%%%%%%%%%%%%%%%%
%%%%%%%%%%%%%%%%%%%%%
%%%%%%%%%%%%%%%%%%%%%
%%%%%%%%%%%%%%%%%%%%%
%%%%%%%%%%%%%%%%%%%%%
%%%%%%%%%%%%%%%%%%%%%
%%%%%%%%%%%%%%%%%%%%%

Next, we discuss the dependence of the phase diagram obtained here on the conduction-band structures
and its consistency with the observed ordering-vector variation in
Ce(Pd$_{1-x}$M$_x$)$_2$Al$_3$. 
The obtained band structures with $\mu=0$
correspond to  the conduction-band structures of CePd$_2$Al$_3$. 
It is difficult to calculate the conduction-band structures of the mixed crystal 
Ce(Pd$_{1-x}$M$_x$)$_2$Al$_3$, and hence
we describe  simply the conduction band of Ce(Pd$_{1-x}$M$_x$)$_2$Al$_3$
as the system to which $2x$ electrons per unit cell are doped, 
by which $\mu$ increases. % 
The calculated phase diagrams %
for $M=\pm 1/2$ and $\pm 3/2$
are plotted versus $\mu$ in Figs.~\ref{figphase} (a) and (b), respectively. 
For $M=\pm 1/2$, 
the $(0,0,1/2)$-AF, F, $(1/2, 0, 1/2)$-AF, $(1/3,1/3,0)$-AF, and $(0,0,1/2)$-AF
 orders are most stable
in the regions $-0.31$ eV $<\mu<$ $-0.09$ eV, $-0.08$ eV $<\mu<$ 0.13 eV, 
0.20 eV $<\mu<$ 0.35 eV, 
0.42 eV $<\mu<$ 0.51 eV, %% 
and 0.56 eV $<\mu<$ 1.03 eV,  respectively. 
For $M=\pm 3/2$, %
the $(0,0,1/2)$-AF, F, $(1/2, 0, 1/2)$-AF, $(1/2,0,0)$-AF, and $(0,0,1/2)$-AF
 orders are most stable
in the regions $-0.31$ eV $<\mu<$ $-0.07$ eV, $-0.06$ eV $<\mu<$ 0.13 eV, %
0.17 eV $<\mu<$ 0.41 eV, 
0.42 eV $<\mu<$ 0.53 eV, 
and 0.56 eV $<\mu<$ 0.99 eV,  respectively. 
The difference in the phase diagram between $M=\pm 1/2$ and $\pm 3/2$ are 
not significant, except for the region $\mu \sim 0.5$ eV, 
suggesting that the difference in the charge distribution anisotropy
between $M=\pm 1/2$ and $3/2$ states is not crucial 
in the determination of the ordering vector, 
compared with the shift of $\mu$ caused by the substitution of atoms. 
We note further that the sequence of
the stable phase is nearly invariant, except for the region $\mu \sim 0.5$ eV, 
even
if $B_{\gamma {\bm{k}}} $ is assumed as constant, 
indicating that the $\bm{k}$ dependence of $B_{\gamma {\bm{k}}} $ 
is not important unless it is too strong.

%%%%%%%%%%%%%%%%%%%%%
\begin{figure}
\includegraphics[width=10.5cm]{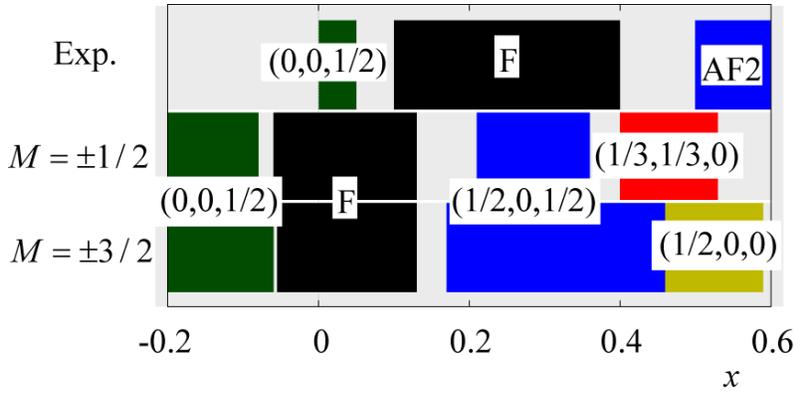} % Here is how to import EPS art
\caption{\label{figx} 
(Color online)
The range of the experimentally\cite{PRB70-174429, PhB359-287, JPCM18-5715}
 (top) and theoretically 
($M=\pm1/2$ in the middle, and $\pm 3/2$ in the bottom) obtained stable ordering vectors  versus
$x$. 
}
\end{figure}
%%%%%%%%%%%%%%%%%%%%%
We plot the range of the stable ordering vectors versus
$x$
for 
$M=\pm1/2$ and $\pm 3/2$, in comparison with the experimental results\cite{PRB70-174429, PhB359-287, JPCM18-5715},  in Fig.~\ref{figx}. 
We obtain the sequence 
of the AF order with $\bm{Q}=(0,0,1/2)$, the F order with $\bm{Q}=(0,0,0)$, 
and another AF order, of which the 
ordering vector is $(1/2,0,1/2)$. 
This sequence is in good agreement with the observed one for 
Ce(Pd$_{1-x}$M$_x$)$_2$Al$_3$ with increasing $x$. 
In order for the theoretical results to correspond with the experimental ones quantitatively, 
it is required to 
shift $x$ by approximately +0.1, which 
corresponds to the change of the conduction-electron number
by 0.2.

To dissolve this discrepancy, it must be allowed to assume
a small change in the conduction-band structures. 
If the 18th conduction band, which are thoroughly unoccupied in our band calculation
as plotted in Fig.~\ref{figdos}, 
are lowered so that it possesses
small electron Fermi surfaces with the filled electron number $\sim 0.2$, 
we may obtain effectively smaller electron filling numbers for the 16th and 17th bands. 
Under this modification in the conduction-band structures, 
we can relate
the $(0,0,1/2)$-AF order for $\mu\sim -0.1$ eV, % 
the F order
in the vicinity of $\mu=0$, and the
$(1/2,0,1/2)$-AF order for $\mu \sim 0.2$ eV %  
to the AF order of CePd$_2$Al$_3$, 
the F order of Ce(Pd$_{1-x}$M$_x$)$_2$Al$_3$ for $0.1<x<0.4$, 
and the AF order  for $x>0.5$, respectively. 
The $(0,0,1/2)$-AF order 
is caused by the RKKY interaction through the structure of the 16th  conduction band, 
while the $(1/2,0,1/2)$-AF order 
is due to the 17th band, and 
the F order  
is caused by the crossover from the 16th to the 17th band. 

Finally, we discuss the justification of the mean-field approximation (MFA). 
In our model, the stable magnetic order is determined mainly by the signs of the RKKY
interaction between the neighboring Ce sites. 
Therefore, the stable ordering vector will not alter substantially
even if the approximation is improved beyond the MFA, but
the transition temperature will be largely changed. 
As a conclusion, the MFA is sufficient inasmuch as the relative stability 
between the various magnetism is concerned. 
A calculation based on an improved approximation is left in a future issue. 

\section{\label{sec4} Conclusion}
In this paper, we have studied the variation of the ordering vectors of  Ce(Pd$_{1-x}$M$_x$)$_2$Al$_3$
on the basis of the anisotropic  RKKY interaction model, using the realistic conduction-band structures
obtained by the band calculation. 
It is understood that 
the conduction-band structures play an important role in determining the ordering vectors, 
in the way similar to the nesting effect in the Hubbard model. 
By treating the substitution of Ag or Cu for Pd atoms as the electron doping to the conduction bands
of CePd$_2$Al$_3$, which are obtained from the band calculation,  
we have obtained the variation of the ordering vectors consistent with experimental
results
for Ce(Pd$_{1-x}$M$_x$)$_2$Al$_3$. 
The $(0,0,1/2)$-AF order of CePd$_2$Al$_3$ is caused by the movement 
of the iso-energy surface of the 16th band to
the central plane of the FBZ with increasing $\varepsilon$. 
On the other hand, the AF order for $x>0.5$ is caused by the movement of the iso-energy 
surface of the 17th band, 
for which the ordering vector is predicted to be $(1/2,0,1/2)$. 
The F order for $0.1<x<0.4$ is 
caused by the crossover of the main conduction band from the 16th to the 17th band. 
Thus, we have obtained the variation of the ordering vectors with increasing $\mu$, which is
consistent with that for Ce(Pd$_{1-x}$M$_x$)$_2$Al$_3$, 
using the anisotropic RKKY model 
with the realistic conduction-band structures. 

\section*{Acknowledgments}
The numerical calculation has been carried out by the sx8 supercomputer in YITP, 
Kyoto University. 
The conduction-band structures have been obtained through the use of the ABINIT code, 
a common project of the Universite Catholique de Louvain, Corning Incorporated, 
and other contributors (URL http://www.abinit.org)\cite{abinit1,abinit2,abinit3}.
It relies on an efficient Fast Fourier Transform algorithm\cite{fftabinit} for the conversion of wavefunctions between real and reciprocal space. 
%\section*{References}

\end{document}